# Analysis of Oscillator Phase-Noise Effects on Self-Interference Cancellation in Full-Duplex OFDM Radio Transceivers

Ville Syrjälä, *Member, IEEE*, Mikko Valkama, *Member, IEEE*, Lauri Anttila, *Member, IEEE*, Taneli Riihonen, *Student Member, IEEE* and Dani Korpi

*Abstract*—This paper addresses the analysis of oscillator phase-noise effects on the self-interference cancellation capability of full-duplex direct-conversion radio transceivers. Closed-form solutions are derived for the power of the residual self-interference stemming from phase noise in two alternative cases of having either independent oscillators or the same oscillator at the transmitter and receiver chains of the full-duplex transceiver. The results show that phase noise has a severe effect on self-interference cancellation in both of the considered cases, and that by using the common oscillator in upconversion and downconversion results in clearly lower residual self-interference levels. The results also show that it is in general vital to use high quality oscillators in full-duplex transceivers, or have some means for phase noise estimation and mitigation in order to suppress its effects. One of the main findings is that in practical scenarios the subcarrier-wise phase-noise spread of the multipath components of the self-interference channel causes most of the residual phase-noise effect when high amounts of self-interference cancellation is desired.

## I. INTRODUCTION

FULL-DUPLEX radio technology is based on a fairly old idea of transmitting and receiving signals simultaneously at the same center-frequency. However, due to massive self-interference (SI) caused by direct coupling of strong transmit signal to the sensitive receiver chain, practical implementations of such radios have not been available until

Manuscript received July 1, 2013 and revised November 29, 2013. This work was supported in part by the Academy of Finland (under the project 259915 "In-Band Full-Duplex MIMO Transmission: A Breakthrough to High-Speed Low-Latency Mobile Networks"), the Finnish Funding Agency for Technology and Innovation (Tekes, under the project "Full-Duplex Cognitive Radio"), the Linz Center of Mechatronics (LCM) in the framework of the Austrian COMET-K2 programme, Japan Society for the Promotion of Science (JSPS) under the Postdoctoral Fellowship program and KAKENHI Grant number 25-03718, and Emil Aaltonen Foundation.

V. Syrjälä, M. Valkama, L. Anttila, and D. Korpi are with the Department of Electronics and Communications Engineering, Tampere University of Technology, P.O. Box 692, 33101 Tampere, Finland.

T. Riihonen is with the Department of Signal Processing and Acoustics, Aalto University School of Electrical Engineering, P.O. Box 13000, 00076 Aalto, Finland.

Corresponding author is V. Syrjälä, e-mail: ville.syrjala@tut.fi.

recently [1], [2], [3], [4], [5]. Such full-duplex radio technology has many benefits over the conventional time-division duplexing (TDD) and frequency-division duplexing (FDD) based communications. When transmission and reception happen at the same time and at the same frequency, spectral efficiency is obviously increasing, and can in theory even be doubled compared to TDD and FDD, given that the SI problem can be solved [1]. Furthermore, from wireless network perspective, the frequency planning gets simpler, since only a single frequency is needed and is shared between uplink and downlink. Another possible advantage is that if the devices in a wireless network have full-duplex capability, they can also sense the traffic in the network during their own transmissions. This can lower the amount of needed medium access and radio link control signalling in the network, therefore improving the maximum throughput of networks, as well as significantly lowering the network delays [3].

Despite of the various benefits, there are, however, still many practical implementation related issues in building commercial small handheld or portable radio devices utilizing full-duplex technology, especially with low-cost deep-submicron integrated circuit technologies. The biggest challenge is the self-interference phenomenon [1], [2], [3], [4], stemming from the imperfect electromagnetic isolation of the transmitter and receiver parts in the overall transceiver. This isolation can be partially assisted by having physically separate transmit and receive antennas, as reported e.g. in [1], [2] and [3], which, depending on the center-frequency and physical separation, yields typical isolations in the order of 20-40 dB or so [1], [2]. The other central element in self-interference suppression is active cancellation, at both RF/analog and digital parts of the receiver chain, using the transmit signal as the reference [1], [2], [3], [4], [6], [7]. The most common analog cancellation approach, like reported, e.g., in [1], [3], [8] and [9], is based on subtracting the actual transmit RF waveform, properly aligned in time, amplitude and phase, from the receiver input. This results in fairly low instrumentation complexity, but can only suppress the dominant SI component while the possible multipath components are then processed in the digital cancellation phase. As an alternative, some recent works [10], [11], [15] have also reported multipath analog/RF cancellation, which increases the RF instrumentation complexity but can in principle then suppress the overall SI, including multipath components, more accurately. Yet another, very interesting alternative is to deploy an additional reference transmitter branch, from digital baseband up to RF, such that an accurate RF cancellation signal can be regenerated. Such works are reported, e.g., in [2] and [5], and can also support multipath cancellation already at RF stage. Such structure does, however, require the additional transmit signal, increasing the overall transceiver complexity. Furthermore, since a separate transmitter chain is used for the RF cancellation signal regeneration, suppressing, e.g., the power amplifier nonlinear distortion occurring in the main transmit path gets potentially more complicated compared to the structures where the main transmit path RF signal is used as reference [1], [3].

While considerable progress in antenna development as well as in analog and digital SI cancellation has been reported in the recent years, the current technology is not yet mature enough, e.g., for full-duplex user equipment (UE) transceivers in 3GPP Long Term Evolution (LTE) networks[1] [16], [17]. Furthermore, when considering low-cost mass-product commercial devices and underlying deep-submicron integrated electronics, all the circuit imperfections related problems, like the phase noise issue considered in this article, are not even fully recognized yet, as the topic is relatively fresh and has been receiving considerable research interest only over the last 3-5 years. Existing works, such as [1], [2], [3], [5] and [11], can be considered state-of-the-art achievements in laboratory scale implementations, but as shown for example in [18] and [19], all the practical implementation limitations are not yet fully understood when commercial low-cost integrated electronics are to be used.

In this article, we address in detail the oscillator phase-noise phenomenon as one of the performance limiting factors in low-cost full-duplex direct-conversion transceivers. In the existing literature, phase noise has been studied in case of full-duplex relays in [18], and for general full-duplex transceivers in [19] and in its very recent extension [20]. However, the analysis in [19] and [20] is mostly limited to narrow-band signal scenarios and a classical small phase-noise assumption is used in the analysis. The work is also mostly stemming and motivated through the findings in selected experimental laboratory equipment, which adds nice connection to practical observations but also partially limits the general applicability of the analysis. Moreover, and perhaps most importantly, the main part of the analysis work in [19] and [20] is focusing on the scenario where *the transmitter and receiver sides have separate oscillators* and thus also separate phase noise processes. This may be a valid scenario in relay type devices, where the receive and transmit entities can be located even on different sides of a building, and hence have completely separate receiver and transmitter hardware. However, in compact full-duplex transceivers of two-way communication systems, sharing the same oscillator between transmitter and receiver of the device is the realistic scenario, especially since the full-duplex device transmits and receives on a single frequency. This is also then the scenario which this article is mostly focusing on, but for generality and comparison purposes, we cover both cases of (*i*) two independent oscillators and (*ii*) common shared oscillator. Furthermore, in the analysis of this paper, no small phase noise assumption is made which adds to the generality of the analysis. Furthermore, no assumption of narrowband signals is made either, but we specifically focus on modern wideband orthogonal frequency division multiplexing (OFDM) based waveforms forming the basis of physical layers of all emerging radio communication systems. Furthermore, the derivations in this article do not have any limitations set by any specific experimental setup. The analysis also includes explicitly the effect of multipath propagation between the transmit and receive antennas, which is shown to have significant contribution to the overall remaining SI due to phase noise, after realistic analog and digital SI cancellation. Also, the subcarrier level interference structure and spectral broadening in the residual self-interference, caused by phase noise [21], is explored in detail as the final demodulation and detection of OFDM waveforms is done in a subcarrier-wise manner. Hence this is emphasized also in the analysis. All the analysis results are also verified with extensive computer simulations in various cases, and the simulations results are carefully analysed.

The rest of this paper is organized as follows. Section II describes in detail the full-duplex self-interference coupling channel between the transmitter and receiver parts of the transceiver, including the effects of antenna isolation, multipath propagation, transmitter and receiver phase noise, and analog and digital SI cancellation. In Section III, stemming from the previous modelling, subcarrier-wise power of the SI is analysed in OFDM-based full-duplex radio at different stages of the receiver path. In Section IV, the derived analytical results are compared with the simulated ones, and the results are analysed. Finally, Section V concludes the work. In the Appendix, details of derivations for the power of the combined frequency-domain phase-noise effect of the transmitter and receiver phase noise processes are given.

---

[1]As a concrete example, we consider Power Class 3 LTE UE with nominal maximum transmit power of 23 dBm [16]. Now, if the UE antenna separation is, e.g., 30 dB, and analog and digital SI cancellation capabilities are say 30 dB and 50 dB, respectively, which represent fairly optimistic values [3], then the remaining self-interference power is still around −87 dBm, when referenced back to the receiver input. This is far from the 3GPP LTE UE receiver reference sensitivity level of −110 dBm, which includes the thermal noise level, interference margin and receiver noise figure utilizing a single resource block mode [16], [17].

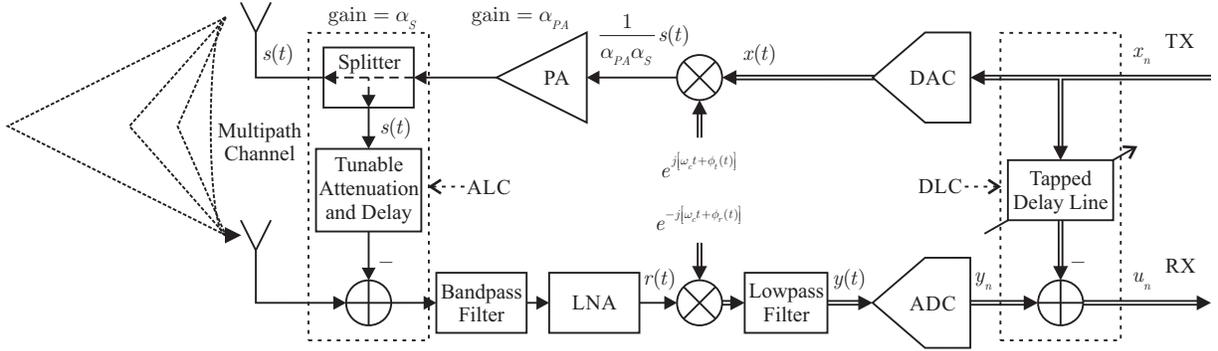

Fig. 1. Principal illustration of the transmitter and receiver analog front-ends of direct-conversion architecture based full-duplex transceiver, including also transmitter and receiver phase noise processes. ALC and DLC refer to analog and digital linear cancellation, respectively.

## II. Full-Duplex Transceiver Principle, Phase Noise and Self-Interference

In this section, the full-duplex transceiver principle is first shortly reviewed, especially from the SI problem point of view. Then, a fundamental signal model is given for the resulting SI with transmitter and receiver oscillator phase-noise processes, including the effects of the multipath channel, antenna isolation, and analog and digital SI cancellation.

### A. Full-Duplex Transceiver Principle

A principal illustration of a generic full-duplex transceiver analog front-end, deploying the direct-conversion radio architecture [3], is given in Fig. 1. As the figure illustrates, separate transmit and receive antennas are deployed, as, e.g., in [1], [2], [3], and hence also assumed in this paper. Analog linear cancellation (ALC) is deployed at the very first stage of the receiver input, to prevent the strong self-interference saturating the whole receiver chain. The needed attenuation and delay of the ALC depend on the characteristics of the main propagation path linking the transmitter and receiver antennas, and on the used transceiver components. Typical reported antenna isolation and ALC numbers are in the order of 20-40 dB and 10-40 dB, respectively [1], [2], [3]. Like was already identified in the Introduction, the ALC principle depicted in Fig. 1 is only one of the many alternatives, but is assumed in this work to facilitate low-cost RF circuit implementations. Further suppression of the SI is then obtained by digital linear cancellation (DLC) inside the transceiver digital front-end. In the DLC, the goal is to estimate the whole SI coupling channel including the multipath components (which are typically ignored in the ALC stage), and then suppress the remaining SI, based on accurate channel estimate and known transmit data ($x_n$ in Fig. 1) [2], [3]. This is done by feeding the digital samples at the transmitter before digital-to-analog conversion (DAC) to a tuneable tapped delay line, which is tuned based on the SI-channel estimates. These samples are then subtracted, properly synchronized, from the signal at the receiver after the analog-to-digital conversion (ADC). In general, mostly linear cancellation solutions have been reported so far in the literature for both ALC and DLC processing interfaces, while recently [8], [11], [12], [13], [14], also nonlinear digital cancellation has been demonstrated.

### B. Self-Interference Model in Full-Duplex Radio after Analog and Digital Cancellation

Now, let us assume that the power amplifier (PA) with amplification factor of $\alpha_{PA} \in \mathbb{R}$ is relatively linear, the splitter gain (attenuation) factor is $\alpha_S \in \mathbb{R}$, and let us denote the complex baseband waveform before the I/Q upconversion at the transmitter by $x(t)$. Practical PAs are typically nonlinear components, but the assumption of linear PA is made here since the focus is on the phase noise induced effects instead of PA. Then, we can write the carrier-modulated RF waveform at the transmitter output as

$$s(t) = \text{Re}\left[\alpha_{PA}\alpha_S x(t) e^{j[\omega_c t + \phi_t(t)]}\right]. \tag{1}$$

Here, $\omega_c = 2\pi f_c$ is the angular oscillator (carrier) frequency and $\phi_t(t)$ is the *phase noise of the oscillator of the transmitter side*. In the individual formulas, time-origin reference is set at the upconverting I/Q mixer interface and the delays in signal propagation and coupling towards receiver side, including the multipath components, are referenced to that. Furthermore, for simplicity and without loss of generality, mixers are assumed unit-gain components.

Next, we assume that the low-noise amplifier (LNA) at the receiver side is relatively linear, and that the total gain and delay between the signal splitter at the transmitter and the ALC combiner element at the receiver, including also the isolation between the transmitter and the receiver antennas (refer to Fig. 1), are $\alpha$ and $\delta$, respectively. Then, the attenuator/delay unit in the ALC processing should ideally be tuned so that the total attenuation and delay of the ALC path matches $\alpha$ and $\delta$, respectively. Naturally, perfect matching is never possible, so let us denote the realized ALC attenuation and delay by $\hat{\alpha}$ and $\hat{\delta}$, respectively. Therefore, the remaining SI signal after ALC processing in the receiver, including also the multipath propagation between the transmit and receive antennas, can in general be written as

$$r(t) = h_B(t) * \left[ \alpha s(t - \delta) - \hat{\alpha} s(t - \hat{\delta}) + \sum_{b=1}^{L} \alpha_b' s(t - \delta_b) \right]. \quad (2)$$

In the above, $*$ denotes convolution, impulse response $h_B(t)$ models the joint linear filtering effect of the receiver bandpass filter and low-noise amplifier, $L$ denotes the number of multipath components between TX and RX antennas, $\alpha_b'$ denotes the attenuation of the $b$ th multipath component and $\delta_b$ denotes the corresponding delay where, as mentioned already earlier, the delays also model the possible delay effects of the transmitter and receiver electronics when transmitter I/Q mixer interface is used as reference. Notice that only the SI signal is considered present in the receiver chain, because the focus in the analysis here is indeed on the SI coupling and cancellation characteristics. Thus, the actual useful received signal is omitted for notational convenience. Then, after the I/Q downconversion and lowpass filtering, whose impulse response is denoted by $h_L(t)$, the complex baseband observation of the SI signal can be written as

$$\begin{aligned} y(t) &= h_L(t) * \left[ r(t) e^{-j[\omega_c t + \phi_r(t)]} \right] \\ &= \alpha x(t - \delta) e^{j[-\omega_c \delta + \phi_t(t - \delta) - \phi_r(t)]} \\ &\quad - \hat{\alpha} x(t - \hat{\delta}) e^{j[-\omega_c \hat{\delta} + \phi_t(t - \hat{\delta}) - \phi_r(t)]} \\ &\quad + \sum_{b=1}^{L} \alpha_b' x(t - \delta_b) e^{j[-\omega_c \delta_b + \phi_t(t - \delta_b) - \phi_r(t)]}, \end{aligned} \quad (3)$$

where $\phi_r(t)$ is now the *oscillator phase noise process at the receiver side*. For notational convenience, the responses of the bandpass and lowpass filters are assumed ideal, and hence do not appear explicitly in the latter form of (3). Notice that for generality, we have denoted the receiver phase noise process in (3) with different random process notation, compared to transmitter side. This later enables us to study the two different scenarios of (*i*) independent oscillators where $\phi_t(t)$ and $\phi_r(t)$ are statistically independent random processes and (*ii*) common shared oscillator where $\phi_r(t) = \phi_t(t)$.

Next, after sampling the baseband observation at instants $t = nT_s$, where $T_s$ is the sampling interval, and when characterizing the physical multipath propagation with corresponding tapped delay line, we can rewrite (3) as

$$\begin{aligned} y_n &= y(nT_s) = \alpha x(nT_s - \delta) e^{j[-\omega_c \delta + \phi_t(nT_s - \delta) - \phi_r(nT_s)]} \\ &\quad - \hat{\alpha} x(nT_s - \hat{\delta}) e^{j[-\omega_c \hat{\delta} + \phi_t(nT_s - \hat{\delta}) - \phi_r(nT_s)]} \\ &\quad + \sum_{b=1}^{P} \alpha_b x\big((n - b)T_s - \delta\big) e^{j[-\omega_c (bT_s + \delta) + \phi_t((n-b)T_s - \delta) - \phi_r(nT_s)]}. \end{aligned} \quad (4)$$

Here, $P$ is the maximum delay of the multipath channel in samples and $\alpha_b$ denotes the complex coefficient of the $b$ th component of the sampled multipath channel model. With substitution $z(t) = x(t - \delta)$, we can write (4) then as

$$\begin{aligned} y_n &= y(nT_s) = \alpha z(nT_s) e^{j[-\omega_c \delta + \phi_t(nT_s - \delta) - \phi_r(nT_s)]} \\ &\quad - \hat{\alpha} z(nT_s + \delta - \hat{\delta}) e^{j[-\omega_c \hat{\delta} + \phi_t(nT_s - \hat{\delta}) - \phi_r(nT_s)]} \\ &\quad + \sum_{b=1}^{P} \alpha_b z\big((n - b)T_s\big) e^{j[-\omega_c (bT_s + \delta) + \phi_t((n-b)T_s - \delta) - \phi_r(nT_s)]}. \end{aligned} \quad (5)$$

Now, with reasonable ALC circuitry, $\delta$ and $\hat{\delta}$ can be assumed to be fairly close to each other, and assuming further that the baseband processing bandwidth is in the range of few tens of MHz, typical to, e.g., mobile cellular radio and wireless local area systems, we can impose approximations of the form $z(nT_s + \delta - \hat{\delta}) \approx z(nT_s)$ and $\phi_t(nT_s - \delta) \approx \phi_t(nT_s - \hat{\delta})$. However, in (5), there are also SI terms in which $\delta$ and $\hat{\delta}$ are essentially multiplied with $\omega_c$ which is typically a very high number (e.g. at 1 GHz). In such SI terms, the small ALC delay error still has clear contribution to the overall signal $y_n$. Therefore, we write (5) finally as

$$\begin{aligned} y_n &= y(nT_s) \approx \alpha_0 z(nT_s) e^{j[-\omega_c \delta + \phi_t(nT_s - \delta) - \phi_r(nT_s)]} \\ &\quad + \sum_{b=1}^{P} \alpha_b z\big((n - b)T_s\big) e^{j[-\omega_c (bT_s + \delta) + \phi_t((n-b)T_s - \delta) - \phi_r(nT_s)]} \\ &= \sum_{b=0}^{P} \alpha_b z\big((n - b)T_s\big) e^{j[-\omega_c (bT_s + \delta) + \phi_t((n-b)T_s - \delta) - \phi_r(nT_s)]}, \end{aligned} \quad (6)$$

where

$$\alpha_0 = \alpha - \hat{\alpha} e^{j\omega_c (\hat{\delta} - \delta)} = \alpha - \big(\alpha - \alpha_e\big) e^{-j\omega_c \delta_e}. \quad (7)$$

This model now includes the essential effects of the amplitude error $\alpha_e = \alpha - \hat{\alpha}$ and delay error $\delta_e = \delta - \hat{\delta}$ in the ALC circuitry, phase noise processes of transmitter and receiver, as well as the multipath propagation between TX and RX antennas, and is used in the forthcoming section for detailed subcarrier-wise SI analysis.

Following the notation in Fig. 1, the remaining SI signal including also DLC can finally be written as

$$u_n = y_n - h_{n,DLC} * x_n. \quad (8)$$

Here, $x_n$ are the samples before transmitter DAC and $h_{n,DLC}$ is the impulse response of the tapped delay line used in DLC.

## III. SUBCARRIER-WISE SELF-INTERFERENCE POWER DUE TO PHASE NOISE IN OFDM FULL-DUPLEX RADIO

In this section, the actual subcarrier-wise SI power with phase noise in the full-duplex transceiver is analysed in closed form,

using the signal models derived in the previous section as the starting point. The study is carried out in two distinct cases, namely (*i*) having fully separate oscillators (thus $\phi_t(t)$ and $\phi_c(t)$ being statistically independent random processes) and (*ii*) having a common shared oscillator for TX and RX (thus $\phi_t(t) = \phi_t(t)$). For both cases, closed-form formulas for subcarrier-wise SI power are derived, and are then used to, e.g., analyse the impact of ALC, multipath propagation and DLC on the residual SI power. Also comparisons between the two different oscillator scenarios are made.

### A. Subcarrier-wise Self-Interference Power before Digital Linear Cancellation

As a starting point, the sampled baseband signal model for the SI with phase noise, and including ALC, is given in (6). In OFDM systems, the signals are discrete Fourier transformed (DFT) at the receiver digital front-end for subcarrier level signal processing. Therefore, we also proceed below with subcarrier-wise signal models by imposing appropriate block-wise DFT operation on the sampled SI signal. In the following, the amount of subcarriers in a single OFDM symbol is denoted by $N$ and this is also assumed, for simplicity, as the DFT size. Taking now the $N$-size DFT of $N$ samples of $y_n$ in (6), and assuming correct synchronization of DFT within cyclic prefix duration, yields

$$
\begin{aligned}
Y_k = \mathrm{DFT}_k\left\{y_n : n = 0, 1, \ldots, N-1\right\} = \\
\sum_{b=0}^{P}\left[\left[\alpha_b Z_k e^{-j\left(\omega_c\left(bT_s + \delta\right) + 2\pi kb/N\right)}\right] \otimes \frac{1}{N}\sum_{n=0}^{N-1} e^{j\left(\phi_t\left((n-b)T_s - \delta\right) - \phi_c\left(nT_s\right)\right)} e^{-j2\pi kn/N}\right],
\end{aligned}
\tag{9}
$$

where relative sample index $n = 0$ corresponds to the first sample inside a time-domain OFDM symbol. Here, $\mathrm{DFT}_k\{\bullet\}$ denotes the $k$ th sample of the DFT of the argument vector, $\otimes$ is circular convolution operator and $Z_k$ is the DFT of $z(nT_s) : n = \{0, 1, 2, \ldots, N-1\}$. Next, all the phase noise terms can be written with the help of a single function

$$
J_k(b, \delta) = \frac{1}{N}\sum_{n=0}^{N-1} e^{j\left(\phi_t\left(nT_s - bT_s - \delta\right) - \phi_c\left(nT_s\right)\right)} e^{-j2\pi kn/N},
\tag{10}
$$

where index $b$ is the number of full-sample delays experienced by the transmitter-induced phase noise. Therefore, we can now rewrite (9) with help of (10) as

$$
\begin{aligned}
Y_k &= \sum_{b=0}^{P}\left[\alpha_b Z_k e^{-j\left(\omega_c\left(bT_s + \delta\right) + 2\pi kb/N\right)}\right] \otimes J_k(b, \delta) \\
&\triangleq \sum_{b=0}^{P}\sum_{l=0}^{N-1}\left[\alpha_b Z_l e^{-j\left(\omega_c\left(bT_s + \delta\right) + 2\pi kb/N\right)}\right] J_{k-l}(b, \delta) \\
&= e^{-j\omega_c\delta}\sum_{l=0}^{N-1} Z_l \sum_{b=0}^{P}\left[\alpha_b e^{-j\left(\omega_c bT_s + 2\pi kb/N\right)}\right] J_{k-l}(b, \delta).
\end{aligned}
\tag{11}
$$

Therefore, the power of SI at an arbitrary subcarrier at DFT output can be defined as

$$
\mathrm{E}\left[\left|Y_k\right|^2\right] = \mathrm{E}\left[\left|e^{-j\omega_c\delta}\sum_{l=0}^{N-1} Z_l \sum_{b=0}^{P}\alpha_b e^{-j\left(\omega_c bT_s + 2\pi kb/N\right)} J_{k-l}(b, \delta)\right|^2\right].
\tag{12}
$$

Here, $\mathrm{E}\left[\bullet\right]$ denotes the statistical expectation operator. Now with the assumptions that $\forall k : Z_k$ are independent of each other, $\forall b : \alpha_b$ are independent of each other following the widely used Bello's wide-sense stationary uncorrelated scattering (WSSUS) model [22], $\mathrm{E}\left[Z_k\right] = 0$ and $\mathrm{E}\left[\alpha_b\right] = 0$, and by denoting $\mathrm{E}\left[\left|Z_k\right|^2\right] \triangleq \sigma_k^2$ and $\mathrm{E}\left[\left|\alpha_b\right|^2\right] \triangleq \sigma_{\alpha,b}^2$, we can rewrite (12) through some straight-forward manipulations into form

$$
\begin{aligned}
\mathrm{E}\left[\left|Y_k\right|^2\right] &= \sum_{b=0}^{N-1}\sigma_l^2 \sum_{b=0}^{P}\mathrm{E}\left[\left|\alpha_b e^{-j\left(\omega_c bT_s + 2\pi kb/N\right)} J_{k-l}(b, \delta)\right|^2\right] \\
&= \sum_{b=0}^{P}\sum_{l=0}^{N-1}\sigma_l^2 \sigma_{\alpha,b}^2 \,\mathrm{E}\left[\left|J_{k-l}(b, \delta)\right|^2\right].
\end{aligned}
\tag{13}
$$

Closed-form expressions for $\mathrm{E}\left[\left|J_k(b, \delta)\right|^2\right]$ are then derived in the Appendix, for both cases of having independent transmitter and receiver oscillators and having the same common oscillator at both sides. In Appendix, the free-running oscillator (FRO) model [26] is assumed to simplify the analysis. Below, (13) is shown in its final form for both of the studied oscillator cases, when the results of Appendix are deployed.

### 1) Independent Oscillators Case:

First, let us consider the case where we have independent oscillators at the transmitter and receiver sides. Then by using (24) derived in the Appendix, the subcarrier-wise SI power in (13) reduces to the form

$$
\begin{aligned}
\mathrm{E}\left[\left|Y_k\right|^2\right] = \frac{1}{N^2}\sum_{b=0}^{P}\sum_{l=0}^{N-1}\Big\{\sigma_l^2 \sigma_{\alpha,b}^2 \times \\
\left[-N + \sum_{n=0}^{N-1} 2(N-n) e^{-4\pi n\beta T_s} \cos\left(2\pi(k-l)n / N\right)\right]\Big\}.
\end{aligned}
\tag{14}
$$

Here, $\beta$ is the 3-dB bandwidth of the used FRO oscillator model. The result obviously depends on the phase noise 3-dB bandwidth and other essential parameters like ALC performance, the number of subcarriers, multipath profile and subcarrier spacing. Numerical illustrations will be given in Section IV.

### 2) Common Oscillator Case:

In the case of a common oscillator feeding both the upconversion and the downconversion, (13) can be, with the help of (25) derived in the Appendix, written as

$$\mathrm{E}\left[\left|Y_k\right|^2\right] = \frac{1}{N^2}\sum_{b=0}^{P}\sum_{l=0}^{N-1}\Big\{\sigma_l^2\sigma_{\alpha,b}^2 \times$$
$$\left[-N+\sum_{n=0}^{b}2(N-n)e^{-4\pi n\beta T_s}\cos\left(2\pi\left(k-l\right)n\,/\,N\right)\right. \tag{15}$$
$$\left.+\sum_{n=b+1}^{N-1}2(N-n)e^{-4\pi b\beta T_s-4\pi\beta\delta}\cos\left(2\pi\left(k-l\right)n\,/\,N\right)\right]\Big\}.$$

Again, the expression is straight-forward depending on the essential system parameters and can be easily evaluated for any arbitrary configuration in terms of amount of phase noise, coupling propagation and OFDM waveform.

### B. Subcarrier-wise Self-Interference Power after Digital Linear Cancellation

In DLC, the signal samples at the transmitter before DAC are fed to a tuneable tapped delay line or other digital filter that tries to mimic the total SI channel from transmitter digital front-end to receiver digital front-end. The output samples are then subtracted from the signal at the receiver after ADC. Since all the phase noise impairments take physically place before the DLC, and the reference signal for the DLC is the pure digital transmit signal, phase noise is basically very troublesome in the DLC process. In general, phase noise has two-fold effect on OFDM waveforms when viewed at subcarrier-level. The first effect is the so-called common-phase-error (CPE) that simply refers to common phase rotation of all subcarrier signals [23], [24]. Notice that in our notations, such CPE can still be included in the effective SI channel, and hence partially mitigated as part of the DLC. How well it is mitigated, depends in general on the performance of DLC, which in turn is directly dependent on the quality of the effective SI channel estimates. This will be quantified soon in an explicit manner. The other fundamental impact of phase noise is the so-called intercarrier-interference (ICI) phenomenon [21], [25], [24] which refers to the spread of the subcarrier energy on top of its adjacent subcarriers. In our signal models in (14) and (15), at a given subcarrier $k$, this is visible as the terms in the inner summation for all $l \neq k$. Such subcarrier spreading in the SI signal cannot be removed by DLC, or any other linear processing without sophisticated phase-noise estimation. This will thus heavily limit the overall achievable SI suppression as will be quantified more explicitly below.

In order to quantify the DLC processing and the resulting SI cancellation performance in detail, we take next both the linear channel estimation errors as well as the subcarrier spreading (ICI) due to phase noise into account. Due to the linear nature of DLC processing, the basic structures of subcarrier-level interference power expressions at the output of DLC still follow those of (14) and (15), as the ICI is structurally already included. Due to channel estimation errors, even the linear terms cannot be, however, suppressed perfectly. Hence, in the power analysis, the joint impact results into the effective linear channel multipath powers $\sigma_{\alpha,b}^2$ being replaced with the effective estimation error power, denoted in the following by $\sigma_{ee}^2$. Based on this, and straight-forward manipulations, the subcarrier-wise SI powers *at the DLC output* can be now written as

$$\mathrm{E}\left[\left|U_k\right|^2\right] = \frac{1}{N^2}\sum_{b=0}^{P}\left\{\sum_{\substack{l=0\\l\neq k}}^{N-1}\left[\sigma_l^2\sigma_{\alpha,b}^2\times\right.\right.$$
$$\left.\left(-N+\sum_{n=0}^{N-1}2\left(N-n\right)e^{-4\pi n\beta T_s}\cos\left(2\pi\left(k-l\right)n\,/\,N\right)\right)\right]+ \tag{16}$$
$$\left.\sigma_k^2\sigma_{ee}^2\left[-N+\sum_{n=0}^{N-1}2\left(N-n\right)e^{-4\pi n\beta T_s}\right]\right\}$$

and

$$\mathrm{E}\left[\left|U_k\right|^2\right] = \frac{1}{N^2}\sum_{b=0}^{P}\left\{\sum_{\substack{l=0\\l\neq k}}^{N-1}\left[\sigma_l^2\sigma_{\alpha,b}^2\times\right.\right.$$
$$\left(-N+\sum_{n=0}^{b}2(N-n)e^{-4\pi n\beta T_s}\cos\left(2\pi\left(k-l\right)n\,/\,N\right)+\right.$$
$$\left.\left.\sum_{n=b+1}^{N-1}2(N-n)e^{-4\pi b\beta T_s-4\pi\beta\delta}\cos\left(2\pi\left(k-l\right)n\,/\,N\right)\right)\right]+$$
$$\left.\sigma_k^2\sigma_{ee}^2\left[\frac{(N-b)^2-3N+b}{2}e^{-4\pi b\beta T_s-4\pi\beta\delta}+\sum_{n=0}^{b}2(N-n)e^{-4\pi n\beta T_s}\right]\right\}, \tag{17}$$

for the cases of independent oscillators and common oscillator, respectively. In these expressions, the power of the main multipath channel component $\sigma_{\alpha,0}^2$ has already the antenna separation and ALC suppression included in it. Thus the ideal ALC and DLC cases, in our terminology, correspond to $\sigma_{\alpha,0}^2 = 0$ and $\sigma_{ee}^2 = 0$. Further insight and details are given in Subsection III.C.

Next, in order to shortly map the channel estimation error variance $\sigma_{ee}^2$ to the corresponding overall DLC suppression, a similar approach as in [1], [2], [3] is taken. Notice that this mapping is defined for the reference case of zero phase noise, since the spectral spreading due to phase noise is explicitly already modelled in (16) and (17). Hence, when we want DLC suppression of $d$ (linear scale amplification/suppression ($0 < d < 1$) factor) in case of no phase noise, then the channel estimation error variance $\sigma_{ee}^2$ for each effective channel impulse response tap is

$$\sigma_{ee}^2 = \frac{d\times a}{P+1}, \tag{18}$$

In the above, $a$ denotes ALC suppression, and it is assumed that the multipath coupling channel has the main component and $P$ other components, and the estimation error variance is assumed identical for all the components for simplicity. Here, as in previous literature [1], [2], [3], the DLC suppression $d$ is defined so that it is the extra overall suppression that DLC offers after ALC has already been implemented. In the numerical examples and illustrations in Section IV, for any given ALC and DLC gains, (18) is used to calculate the channel estimation error variances. Notice, again, that in this terminology, the DLC gain refers to achievable digital SI power suppression with zero phase, while the impacts of phase noise are then explicitly built in to (16) and (17) through SI spectral broadening (ICI).

### C. Further Insight on ALC and Multipath Propagation

In the above derivations, antenna separation is already taken into account in the definition of $\alpha$ in (2) and the ALC suppression factor is taken into account in $\alpha_0$ in (7) (so that they do not affect the multipath components). However, it was still left open what is the exact relationship between the amount of antenna separation $c$ (linear scale amplification/suppression $(0 < c < 1)$ factor), ALC-suppression factor and the value of $\alpha_0$. This is addressed below.

Antenna separation $c$ is taken into account only in the main multipath component of the channel, because the separation does exactly that, namely attenuates the main multipath component, because any reflections coming from further away cannot be essentially attenuated with antenna separation in small full-duplex transceivers. However, usually in the literature [1], [2], [3], [4], and thus also in this article, the term *antenna separation* is used to denote *the isolation of the whole SI signal with all multipath components included*. Therefore, if we assume antenna separation of $c$, the suppression of the main component must be more than $c$ so that the whole signal is isolated with the given antenna separation $c$ when also multipath components are present. The used suppression factor for the main signal component $\alpha_0$ is therefore

$$c' = c + (c-1)\sum_{b=1}^{P}\sigma_{\alpha,b}^2, \qquad (19)$$

where it is assumed that the main component of the channel has no attenuation before the isolation factor $c$ is applied and that the powers of the other multipath components $\sigma_{\alpha,b}^2, b \in \{1, 2, \dots, P\}$ are normalized so that direct lossless coupling corresponds to unit power.

Also the actual ALC is then modelled by suppressing only the main multipath component of the channel, as already mentioned after (6) and illustrated in Fig. 1. Typically, the ALC suppression performance is given, similarly as antenna separation, as its capability to suppress the whole SI signal

(including all the multipath components), even though it only suppresses the main component. Therefore, if we desire ALC suppression of $a$ (linear scale amplification/suppression $(0 < a < 1)$ factor), we actually need to suppress the main component as much as

$$a' = \frac{a\sigma_{h,0}^2 + (a-1)\sum_{b=1}^{P}\sigma_{\alpha,b}^2}{\sigma_{h,0}^2}. \qquad (20)$$

Here, $\sigma_{h,0}^2$ is the power of the main component of the channel with antenna separation taken into account. This suppression factor $a'$ is needed when using (14), (15), (16) and (17).

For obvious reasons, both suppression factors also need to be non-negative. Negative values indicate that the desired $c$ or $a$ are simply not achievable in the considered multipath coupling channel, and therefore the maximum achievable $c$ and $a$ are logically dependent on the coupling channel as

$$c > \frac{\sum_{b=1}^{P}\sigma_{\alpha,b}^2}{1+\sum_{b=1}^{P}\sigma_{\alpha,b}^2}, \qquad a > \frac{\sum_{b=1}^{P}\sigma_{\alpha,b}^2}{\sigma_{h,0}^2 + \sum_{b=1}^{P}\sigma_{\alpha,b}^2}. \qquad (21)$$

These are the maximum attainable antenna separation and ALC suppression factor, respectively.

As a final note, we wish to acknowledge again that analog/RF cancellation schemes with more than one paths are basically also possible, and demonstrated e.g. in [11]. In this article, however, as the focus is on low-cost small commercial devices with as simple RF parts as possible, the single-path ALC concept has been deployed. It is also the most common approach in the existing literature and demonstrations.

### IV. SIMULATION SCENARIOS, RESULTS AND ANALYSIS

In this section, the validity of the above analysis results is verified with full waveform simulations of a complete OFDM full-duplex transceiver. First, the simulator is shortly described, followed by numerical specifications of the studied scenarios. Then, the simulation results are given together with the corresponding analytical results. The results are also compared, analysed, and discussed in detail.

### A. Simulator Description

For verification of the analytical results, the SI link of the full-duplex transceiver is simulated as follows. First, transmitter OFDM baseband waveform with 1024 subcarriers, of which 300 on the both sides of the DC-bin are active and carry randomly-drawn 16QAM subcarrier data, is created. The subcarrier spacing is 15 kHz. After the basic waveform generation, a cyclic prefix of 63 samples is added to the signal. The used signal resembles closely the 3GPP LTE [29] downlink signal, and it was thus selected as an example waveform with practical basis and relevance. The given analysis results are, however, valid for arbitrary OFDM signals. After adding the cyclic prefix, the phase noise in the upconversion is modelled into the signal by using the FRO phase-noise model [26]. An RF carrier frequency of 1.875 GHz is assumed as a practical example of typical cellular radio frequencies. These waveform parameters are also summarized in Table I. The signal is then propagating through a multipath coupling channel that is described in more details later. The multipath coupling channel has also the separation between the transmit antenna and the receiver antenna included in it, as described in the previous section. Then at the receiver, ALC is modelled so that the case-dependent desired attenuation for the main multipath component is attained by having appropriate amplitude and phase errors included in the ALC whose values depend directly on the assumed ALC suppression factor. The ALC suppresses only the main multipath component, as in the analysis. After ALC modelling, the FRO based phase noise in the downconversion is modelled. In the common oscillator case, the same phase noise realization is used as in the transmitter part but with delay $\delta$, while in the independent oscillators case, the two realizations are drawn independently. Finally, DLC is modelled so that an appropriate estimation error is assumed in the estimation of the effective multipath coupling channel, as described in the previous section. The digital signal at the transmitter is then processed with the coupling channel estimate, with estimation error included, and the output is subtracted from the received digital signal. In the end, the signal is fed to receiver FFT and the remaining SI power is evaluated numerically. The whole process is repeated over 1000 independent trials of the underlying data and phase noise realizations, to collect reliable statistics.

### B. Parameters for Numerical Results

In all the simulations, the reference subcarrier-level average power at the receiver FFT output is set to 0 dB, so the SI power is given in dB's in relation to that. Therefore the given SI power is also directly the total ALC+DLC suppression in the presence of phase noise. We assume a multipath coupling channel with power profile of $-30$ dB, $-65$ dB, $-70$ dB and $-75$ dB for delays of 0, 1, 2 and 4 samples ($P = 4$), respectively, which is modified from [27] to fit better to full-duplex transceiver scenarios. The $-30$ dB attenuation ($c = 10^{-3}$) of the main tap results from the 30 dB (or 30.00006 dB to be exact) antenna separation, which corresponds to a distance of roughly 20 cm between the antennas [28]. The corresponding delay is $\delta \approx 6.6713 \cdot 10^{-10}$ s. These are found realistic for small handheld/portable devices [28]. These are also close to the values measured in [30] for full-duplex relays, but with a lower, more practical, antenna separation.

TABLE I. PARAMETERS OF THE SIMULATED TRANSMIT WAVEFORM

| Parameter | Value |
|---|---|
| Number of subcarriers | 1024 |
| Number of active subcarriers | 600 |
| Subcarrier modulation | 16QAM |
| Sampling frequency | 15.36 MHz |
| Subcarrier spacing | 15 kHz |
| Cyclic prefix length | 63 samples |
| Carrier frequency | 1.875 GHz |

To study the effects of phase noise on digital and analog SI cancellation, we define two basic scenarios. In the so called "Practical" case, the suppression by ALC (with no phase noise) is assumed to be 30 dB ($a = 10^{-3}$) and the suppression by DLC (with no phase noise) is assumed to be 50 dB ($d = 10^{-5}$), hence the total ALC+DLC suppression in the phase noise free case would be 80 dB. How much phase noise then impacts the total SI suppression is illustrated in the performance figures. The values of 30 dB and 50 dB are chosen since they are close to the reported achievable values in [3] and [28]. This way there is clear connection to recently reported work, though the values are perhaps slightly optimistic due to laboratory scale equipment used in [3] and [28]. In the other scenario, denoted as the "Ideal" case, we use otherwise the same parameters but ALC is now assumed to be ideal, which means perfect suppression of the main SI component in the phase noise free case ($\sigma_{\alpha,0}^2 = 0$). Furthermore, in this "Ideal" case, perfect DLC is also assumed ($\sigma_{ee}^2 = 0$), implying that the DLC suppression, and hence the total SI suppression, would be $\infty$ dB without phase noise. How much phase noise then impacts the total SI suppression is again illustrated in the performance figures. With this "Ideal" case, we can truly push the limits in the SI cancellation process set by the phase noise alone.

The basic descriptions of the "Practical" and "Ideal" cases are also shortly summarized in Table II. In addition to these two basic scenarios, additional performance studies are also carried out where either the ALC or the DLC suppression is varied. These are clearly indicated in the corresponding result figures when applicable. We also wish to note that even though the FRO model is used in the analysis and simulations, the results are fairly generally applicable. This is because when the CPE is seen as part of the linear SI coupling channel, the effective remaining phase noise indeed closely resembles the practical phase-locked loop (PLL) –based oscillators when it comes to inband effects [21], [25] which is the focus in the full-duplex context. More far away noise floor of practical oscillators contribute in practice, e.g., to adjacent channel interference, but this does not impact the SI cancellation and is thus out of the scope of this article. Examples of the effective phase noise realizations are given for reference in Fig. 2 for 50-Hz and 10-Hz phase noise 3-dB bandwidths $\beta$.

TABLE II. ASSUMED REFERENCE ALC, DLC AND TOTAL SI SUPPRESSION VALUES FOR "PRACTICAL" AND "IDEAL" CASES WITHOUT PHASE NOISE. IN CASES WHERE ALC OR DLC VALUE IS FURTHER VARIED, THE FIXED VALUE IN THE TABLE IS REPLACED WITH THE VARIED VALUE

|  | ALC suppression without phase noise | DLC suppression without phase noise | Total suppression without phase noise |
|---|---|---|---|
| Practical Case | 30 dB | 50 dB | 80 dB |
| Ideal Case | Max. attainable, perfect for main SI component | $\infty$ dB | $\infty$ dB |

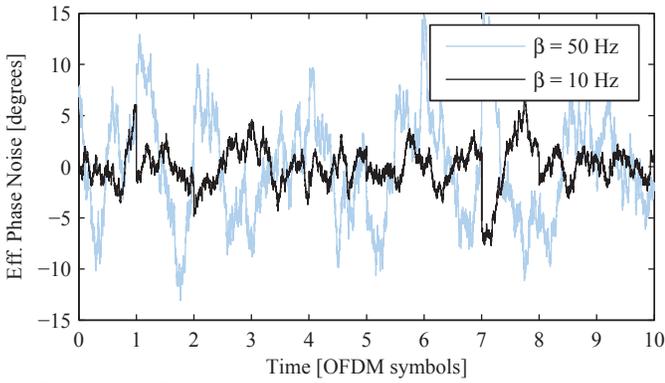

Fig. 2. Example effective phase noise realizations for 50 Hz and 10 Hz phase-noise 3-dB bandwidths ($\beta$) with common phase error removed.

### C. Results and Analysis

#### 1) Principal Spectral Illustrations:
Here, the results corresponding to the previously given scenarios are illustrated and compared with the analytical results. The first results are given in Fig. 3 for the specified Practical case with an example FRO 3-dB bandwidth of $\beta = 50$ Hz. In the common oscillator scenario, since most of the phase noise is cancelled by the downconverting oscillator, the SI level is not far from the level of the SI without phase noise (at −80 dB level in Practical case). In this example, Phase noise causes a noise floor increase of around 4 to 5 dB. In the case of independent oscillators, on the other hand, the phase noise effect is much more severe, as expected intuitively. The SI signal is only 48 dB under its original power. Phase noise thus heavily limits the performance of the DLC, and results in highly problematic receiver scenario. In general, as the figure illustrates, the analytical and simulated results match perfectly.

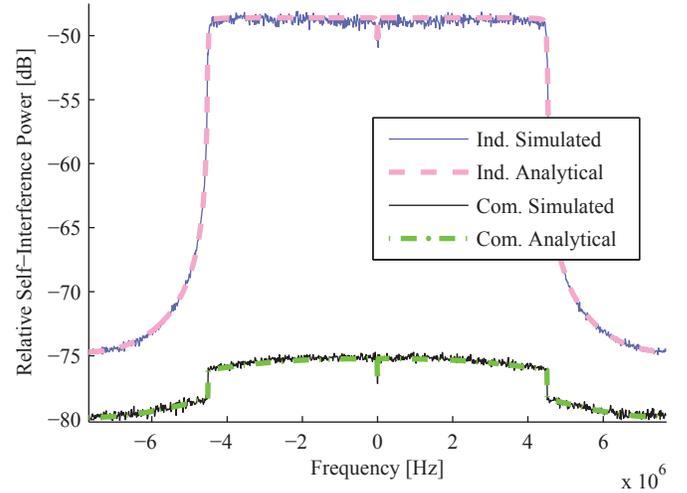

Fig. 3. Relative SI powers at DLC output in the Practical case (defined in Table II) with either two independent (Ind.) FROs or the common (Com.) FRO with 3-dB bandwidth of 50 Hz. Without phase noise, the total SI cancellation would be 80 dB.

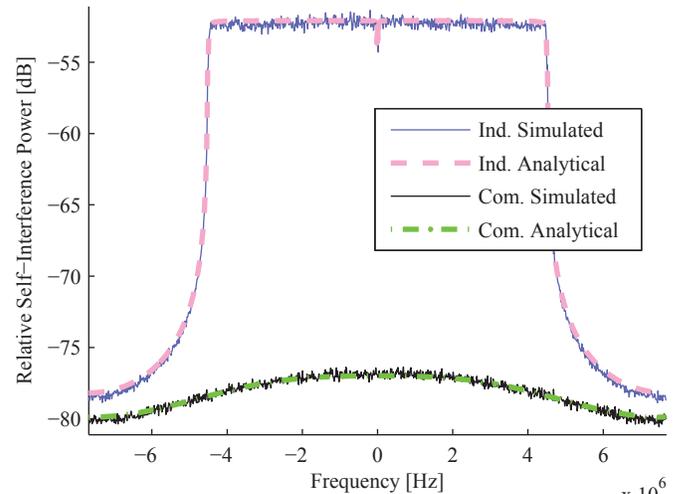

Fig. 4. Relative SI powers at DLC output in the Ideal case (defined in Table II) with either two independent (Ind.) FROs or the common (Com.) FRO with 3-dB bandwidth of 50 Hz. Without phase noise, the total SI cancellation would be 80 dB.

In Fig. 4, the corresponding results for the Ideal case are given for the same FRO with $\beta = 50$ Hz. As in the Practical case, in the case of two independent oscillators, the power of the remaining SI is so high that it significantly limits the performance of the DLC, and thus the whole device. On the other hand, in the common oscillator case, we can see that the theoretical limit for inband SI suppression, in terms of linear SI channel knowledge, is at around −77 dB. This shows that phase noise causes relatively high performance floor for SI cancellation, despite of perfect linear SI channel knowledge, as in this Ideal case, perfect SI cancellation would be obtained without phase noise. So, in principle, if very high SI suppression levels are required, while using low-cost oscillators, it is most likely vital to have some means of phase noise estimation and mitigation implemented in the receiver path. In general, the full spectral illustrations with snap-shot parameter values in Fig. 3 and Fig. 4 are given so that the reader can easily see that the analytical results match practically perfectly with the simulated results subcarrier by subcarrier. In the following, we mostly then focus on showing the SI power values as functions of various elementary parameters, instead of spectral illustrations.

*2) Effect of Varying Phase Noise Levels:* The results for average relative remaining inband SI power at DLC output as a function of $\beta$, ranging from 0 Hz to 1 kHz, are given in Fig. 5 for the Practical and Ideal cases, respectively. In the Practical case with the independent oscillators, one can see that the phase noise causes the SI powers to increase very fast, and even when $\beta$ is only at a nominal value of around 1 Hz, remaining SI level is already more than −65 dB. In the common oscillator case, the interference due to phase noise starts to rise quite steadily after around $\beta = 5$ Hz (denoted by a vertical dot line in the figure). At $\beta = 1$ kHz the interference has already caused a 15 dB decrease in the achieved SI cancellation. In the Ideal case, the results are somewhat different. Using different oscillators increases the SI level around 30 dB compared to the common oscillator case. In the common oscillator case, the interference level rises quite fast to −85 dB at around $\beta = 10$ Hz. This implies that even with small phase noise, phase noise estimation and mitigation would be very useful depending on the desired total SI suppression, if ALC and DLC otherwise work well. For example, in an LTE UE transceiver, the required total SI suppression (including antenna separation, ALC and DLC) is up to 133 dB (from maximum 23 dBm transmit power to −110 dBm, when referenced to receiver input, including thermal noise level of one resource block, interference margin and noise figure of UE receiver). In order to have the phase noise degradation staying below the noise plus interference level, if a practical number of 30 dB antenna separation is assumed (and therefore the needed ALC+DLC suppression is around 103 dB), effectively only $\beta < 0.1$ Hz phase noise levels are tolerated, even in the common oscillator case. Notice also the difference in Fig. 5 between the Practical and Ideal cases with independent oscillators when phase noise rises to high levels. The difference is explained by the difference between the ALC performances, because in that region the ALC performance dictates the overall performance (as heavy phase noise greatly impairs the DLC suppression). In Ideal case, the ALC performs a bit better (the maximum attainable ALC suppression is a bit higher than the ALC suppression of the Practical case). In the same oscillators case, the curves are overlapping at high phase noise levels, because the ALC and downconversion suppress the phase noise in both cases evenly.

*3) Effect of Varying Multipath Powers:* In Fig. 6, the results are given as a function of relative changes in the coupling channel multipath profile. The same channel power profile is used as above (with antenna separation of 30 dB, and relative to that 1st tap is 0 dB, 2nd tap is −35 dB, 3rd tap is −40 dB and 5th tap is −45 dB for delays of 0, 1, 2 and 4 samples, respectively) as the baseline, but the powers of all the taps other than the first one are varied according to the amount of decibels denoted by the horizontal axis. This models weakening or strengthening the reflecting multipath components by the given dB amount. For the Ideal case, the difference between the independent and the common oscillator cases remains unchanged even when the channel conditions are varied. This is because with perfect ALC, the remaining phase noise effects are stemming only from the non-main multipath components, and thus changing their relative power changes the phase noise impacts similarly in both of the cases.

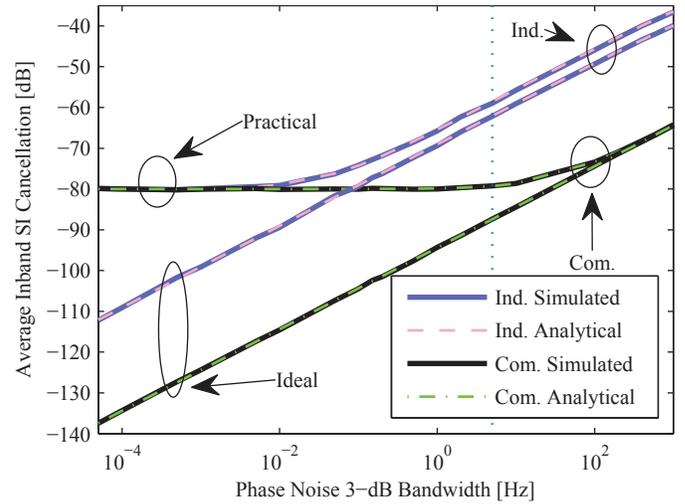

Fig. 5. Average inband SI cancellation at DLC output in Practical and Ideal cases (defined in Table II) as a function of 3-dB bandwidth of oscillator phase noise generated by either two independent (Ind.) FROs or the common (Com.) FRO. Without phase noise, the total SI cancellation would be 80dB (Practical) or $\infty$ dB (Ideal). Vertical dot line marks the phase-noise 3-dB bandwidth of 5 Hz for reference.

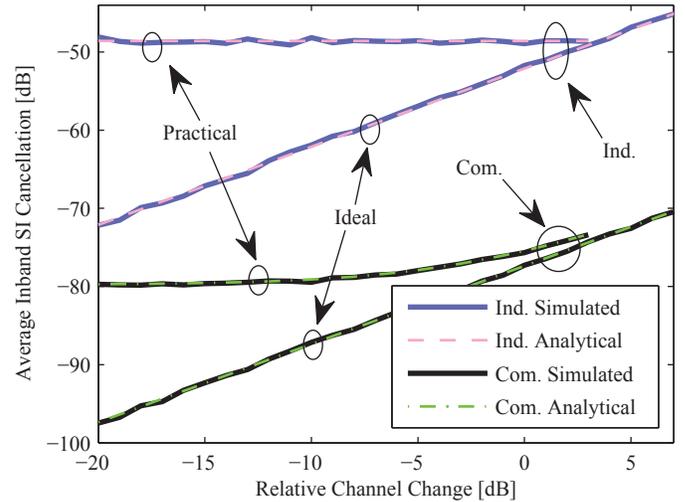

Fig. 6. Average inband SI cancellation at DLC output in Practical and Ideal cases (defined in Table II) as a function of the relative channel change with fixed phase-noise 3-dB bandwidth of 50 Hz generated by either two independent (Ind.) or the common (Com.) FROs. Without phase noise, the total SI cancellation would be 80dB (Practical) or $\infty$ dB (Ideal).

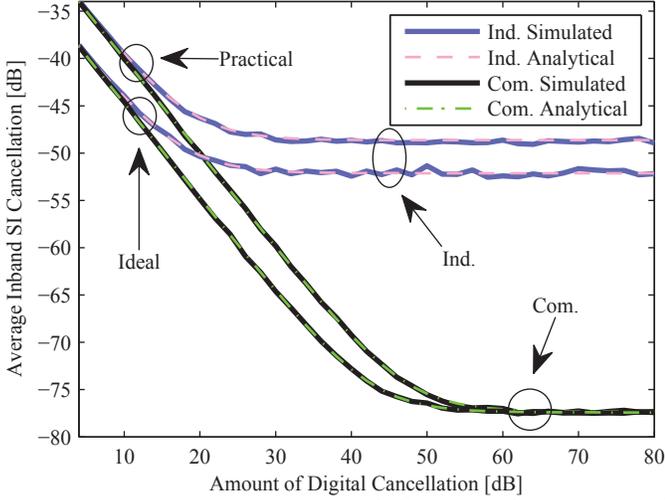

Fig. 7. Average inband SI cancellation at DLC output in Practical and Ideal cases (defined in Table II) as a function of the amount of digital cancellation with fixed phase-noise 3-dB bandwidth of 50 Hz generated by either two independent (Ind.) FROs or the common (Com.) FRO. Without phase noise, the total SI cancellation would be 80dB (Practical) or $\infty$ dB (Ideal).

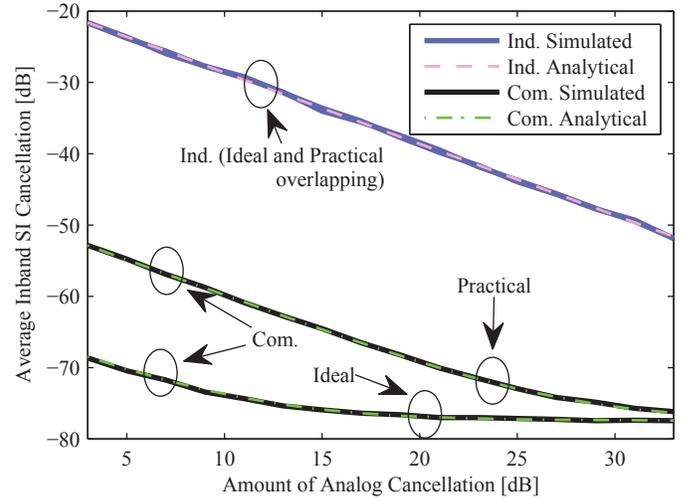

Fig. 8. Average inband SI cancellation at DLC output in Practical and Ideal cases (defined in Table II) as a function of the amount of analog cancellation with fixed phase-noise 3-dB bandwidth of 50 Hz generated by either two independent (Ind.) FROs or the common (Com.) FRO. Without phase noise, the total SI cancellation would be 80dB (Practical) or $\infty$ dB (Ideal).

In the Practical case with independent oscillators, the phase noise from the main multipath component limits the performance, so the curves are essentially straight horizontal lines. In the corresponding curve in the common oscillator case, one can see how the stronger multipath components start to limit the performance already at low relative power levels. In the Ideal case, we can see a linear increase of the average SI power as the multipath components get relatively stronger. Note that in the Practical case, curves end at around 3 dB relative channel change point, because after that it is not possible to attain the desired 30 dB ALC due to too strong multipath components. The curves corresponding to the Ideal case, however, continue forward, because in the Ideal case the best possible ALC is always used independently of it being less or more than ALC in the Practical case.

*4) Effect of Varying DLC Gain:* In Fig. 7, the amount of digital cancellation is varied while ALC performance is fixed. In the "Ideal" case, non-ideal DLC is thus now coupled with the ideal ALC. From these results, we are able to see how the performance of DLC limits the system performance. In the Ideal case, when DLC is non-ideal, the SI levels decrease as a linear function of the DLC performance until at some point it floors to an error floor set by the phase noise of the multipath components. In the case of different oscillators, the performance starts to clearly deviate from the linear curve at 15 dB of DLC. This implies that in such a setup, already relatively low phase noise ($\beta = 50$ Hz) begins to affect the SI cancellation performance when DLC is set to fairly low level of 15 dB. In the common oscillator case, on the other hand, flooring starts to show at around 40 dB of DLC. This value is around the best values reported in current literature, but still very low compared to what is needed in order to implement full-duplex transceivers that comply with modern mobile communications standards, such as 3GPP LTE. With the used coupling channel, the maximum attainable ALC with $\sigma_{\alpha,0}^2 = 0$ is around 33.5 dB. This explains why the Ideal case curves are so close to the Practical case ones, with the exception of the flooring of the independent oscillators case, where the ALC and therefore the main multipath component limit the SI cancellation performance.

*5) Effect of Varying ALC Gain:* In Fig. 8 the amount of analog cancellation is in turn varied. In the Ideal case, a non-ideal ALC is thus now coupled with the ideal DLC. One can see that the non-ideal ALC limits heavily the system performance in the case of independent oscillators, and therefore the curves with independent oscillators linearly decrease as a function of the amount of the ALC. With these parameters, if ALC is increased even more, it would also floor to the noise level seen in the curves of the case with the common oscillator. In the Practical case, increasing ALC performance increases the SI removal until it at some point reaches the flooring level set by the phase noise. The flooring is caused by the phase noise in the multipath components since ALC obviously only affects the main component. The same is seen in the Ideal case, but the curve starts from a slightly lower level because of perfect DLC. After certain level of ALC it does not anymore benefit to have better DLC, because the phase noise sets the performance limit.

*6) Effect of Varying TX-RX Delay:* The delay that the signal experiences from the upconverting mixer at the transmitter to the downconverting mixer at the receiver is varied in Fig. 9. The delay is varied from around 0.67 ns to around 65 ns (delay of one sample). If mapped to corresponding distances, these correspond to from around 20 cm to around 19.5 m distances between the oscillator interfaces. With higher delays, this is thus quite a theoretical analysis because the varying delay between oscillators should also affect the electrical antenna separation and isolation, through increasing propagation losses already. However, as the exact mapping of the varying distance to the varying antenna isolation depends on, e.g., the antenna and other transceiver implementation details, we first ignore this effect and study how the delay alone affects the performance. This is in any case interesting, especially in the common oscillator case, as the delay intuitively has an impact on the phase noise self-cancellation in the downconversion process, especially for the multipaths but also the main path if ALC is not perfect. More specifically, in the case of independent oscillators, the SI power curves vs. distance/delay would just be straight lines since the delay does not affect the statistical properties of the combined phase noise in any way, as also implied by (14). That is why the curves corresponding to the independent oscillators case are left out and the numbers are simply reported here in the text. In that case, the average inband SI cancellation levels are −48.6 dB and −52.1 dB for Practical and Ideal cases, respectively. In the common oscillator case, however, illustrated in Fig. 9, we see that both the Practical and the Ideal case curves logarithmically increase as a function of the delay between the upconverting and downconverting oscillators. These are natural results as the delay between the oscillators directly affects the phase noise effects generated by all the multipath components, and in particular the phase noise self-cancellation in the downconversion process, i.e., as the delays of the components increase, the phase noise self-cancellation deteriorates.

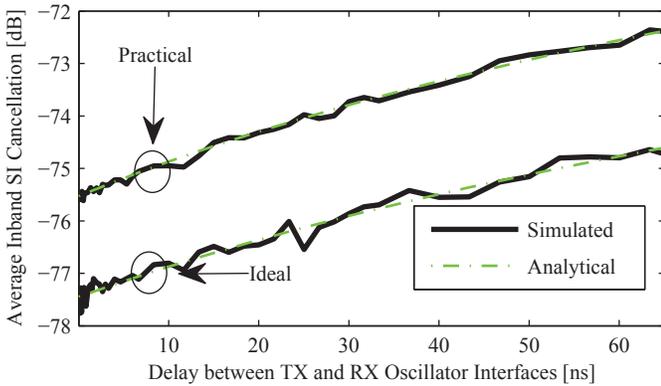

Fig. 9. Average inband SI cancellation at DLC output in Practical and Ideal cases (defined in Table II) as a function of the delay between the TX and RX oscillator interfaces with fixed phase-noise 3-dB bandwidth of 50 Hz generated by the common FRO. Without phase noise, the total SI cancellation would be 80dB (Practical) or $\infty$ dB (Ideal).

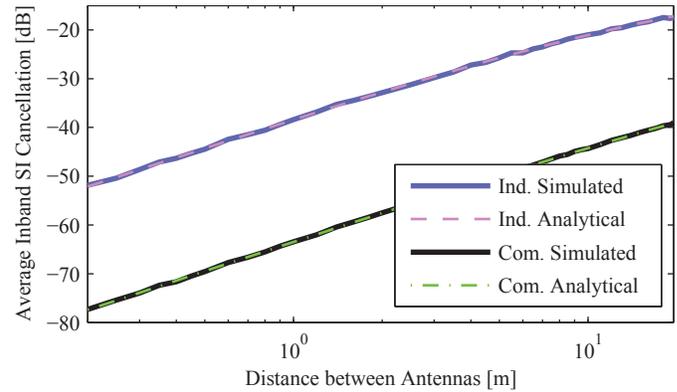

Fig. 10. Average inband SI cancellation at DLC output in the Ideal case (defined in Table II) as a function of the distance between the antennas with fixed phase-noise 3-dB bandwidth of 50 Hz generated by the common FRO. Without phase noise, the total SI cancellation would be $\infty$ dB (Ideal).

*7) Effect of Varying TX-RX Distance:* In Fig. 10, the obtained results are given with different distances between TX and RX antennas such that also the propagation losses are explicitly taken into account. The same multipath channel scenario, as previously, is used as the starting point, and 30-dB antenna separation is assumed with the default distance of 20 cm. As the distance is increasing, excess attenuation is then modelled to all multipath components, in addition to increasing delays. The excess attenuation is directly calculated by adding extra attenuation to each multipath component according to the widely used free-space propagation model for far-field communications [31]. In general, the used rather simple propagation model is a compromise of being able to assess and demonstrate the device performance with different distances and relatively realistic and easily parameterizable loss model. We also emphasize that as the distance increases to 10 to 20 m or so, the power delay profile of the received multipath components already essentially resembles the well-established short-range or small-cell communications propagation models reported e.g. in [29]. This is logical as when the transmitter-receiver distance is already in the order of 10-20m, the propagation conditions more and more resemble an ordinary radio link.

The obtained results in Fig. 10 are only depicted in the ideal case. This is because the effective multipath channel is varying for varying transmitter-receiver distance, and thus the achievable ALC gain also changes with varying distance and hence fixed ALC gain is not a feasible assumption. The results in Fig. 10 show that the distance between the antennas has a huge effect on the overall ALC+DLC performance in both the independent oscillators and the common oscillator cases. In both cases, the SI cancellation performance degradation is explained by the fact that the main multipath component whose cancellation is pursued in the ALC processing is getting relatively weaker and weaker compared to the other multipath components as the distance increases. Since the ALC always inherently mitigates the phase noise effect from the main multipath component, while the other multipath components get relatively more and more powerful, the overall phase noise effect on the SI cancellation gets worse and worse. The performance difference between the independent oscillators and the common oscillator cases starts from around 26 dB at 20 cm distance and ends to around 22 dB at around 20 meters distance (corresponds to one sample delay). This is natural, since the inherent phase noise self-cancellation in the downconversion suffers more and more in the common oscillator case when the delay increases. The overall performance loss due to phase noise in the SI cancellation is natural when the relative dominance of the main coupling component gets smaller. However, when the distance increases, it also means that the natural isolation between the antennas gets higher and higher. Therefore, even though the phase noise effects get heavier from the SI cancellation perspective as the distance increases, the overall full-duplex transceiver performance is increasing. With the natural isolation by the channel, ALC and DLC all taken into account, the total suppressions of the SI signal for the independent and the common oscillator cases are 82 dB and 108 dB, respectively, for 20 cm distance, and 88 dB and 110 dB, respectively, for 20 m distance. Overall, this study shows that with higher distances and especially delays between transmitter and receiver chains, the phase noise effects are emphasized, which thus motivates for either better oscillator optimization or development of explicit phase noise estimation and suppression methods.

The results also indicate that especially with strong multipaths and long coupling delays, improving the ALC performance and in particular its capability to process multipath components, is essential when operating with practical oscillators with considerable phase noise.

## V. Conclusions

In this article, the impacts of phase noise in upconverting and downconverting oscillators of full-duplex direct-conversion OFDM radio were analysed in detail. The analysis takes into account realistic isolation and multipath propagation between transmitting and receiving antennas, analog/RF self-interference cancellation and digital self-interference cancellation. Under these assumptions, closed-form expressions were derived for the remaining subcarrier-wise self-interference power at the receiver path, covering both cases of having either independent oscillators or a single common oscillator for the upconversion and downconversion. The analysis takes explicitly into account the spectral broadening in the self-interference signal, caused by phase noise.

A general outcome of the analysis is that phase noise can seriously compromise the self-interference cancellation in the receiver path, especially in the case of independent oscillators. Hence it can be concluded that in a full duplex transceiver, it is beneficial to use the same common oscillator in the upconversion and downconversion. Furthermore, if high amounts of SI cancellation are pursued, the oscillator should be of very high quality, or some form of phase noise estimation and mitigation should be built into the self-interference cancellation processing in the receiver path. It was also shown that phase noise limits especially the performance of the digital cancellation of the self-interference, even when phase noise level is relatively low, since the digital cancellation samples do not contain, by default, any reference to phase noise. Another key observation is that after the RF/analog cancellation, the phase noise in the self-interference multipath components becomes a limiting factor in self-interference cancellation. Hence the scenarios with strong reflections and long delays are generally shown to be most problematic.

## Appendix

## Derivation of $\mathrm{E}\left[\left|J_k(b,\delta)\right|^2\right]$

In this appendix, we derive the expression for the power of $J_k(b,\delta)$ in (10), i.e., $\mathrm{E}\left[\left|J_k(b,\delta)\right|^2\right]$, following partly the derivations in [33]. This is done for the two cases of having either independent oscillators or the common oscillator in the transmitter and the receiver paths. To begin with, we denote the power with $P_k$ and define it as

$$
\begin{aligned}
P_k &= \mathrm{E}\left[\left|J_k(b,\delta)\right|^2\right] = \mathrm{E}\left[\left|\frac{1}{N}\sum_{n=0}^{N-1} e^{j\left[\phi_t(nT_s-bT_s-\delta)-\phi_r(nT_s)-2\pi kn/N\right]}\right|^2\right] \\
&= \frac{1}{N^2}\mathrm{E}\left[\left(\sum_{n=0}^{N-1} e^{j\left[\phi_t(nT_s-bT_s-\delta)-\phi_r(nT_s)-2\pi kn/N\right]}\right)\times \right. \\
&\qquad \left. \left(\sum_{n=0}^{N-1} e^{-j\left[\phi_t(nT_s-bT_s-\delta)-\phi_r(nT_s)-2\pi kn/N\right]}\right)\right].
\end{aligned}
\tag{22}
$$

Next, we assume that the phase noise process is Brownian motion, namely Wiener process, which on the other hand means that the phase noise is generated by a free-running oscillator. This assumption makes the analysis tractable, but still also generally fairly applicable, because previous studies have shown that free-running oscillators with Brownian motion phase noise give in the end similar phase noise characteristics as, e.g., phase-locked loop based oscillators [21], when the common phase error is compensated for, as is done later in the analysis. With such free-running phase-noise assumption, (22) can be next written as

$$
\begin{aligned}
P_k &= \frac{1}{N^2}\sum_{n=0}^{N-1}\sum_{n'=0}^{N-1}\mathrm{E}\left[e^{j\left[\phi_t(nT_s-bT_s-\delta)-\phi_t(n'T_s-bT_s-\delta)+\phi_r(n'T_s)-\phi_r(nT_s)+2\pi k(n'-n)/N\right]}\right] \\
&= \frac{1}{N^2}\sum_{n=0}^{N-1}\sum_{n'=0}^{N-1} e^{-\frac{1}{2}\mathrm{E}\left[\left|\phi_t((n-b)T_s-\delta)-\phi_t((n'-b)T_s-\delta)+\phi_r(n'T_s)-\phi_r(nT_s)\right|^2\right]+j2\pi k(n'-n)/N}.
\end{aligned}
\tag{23}
$$

The above expression is based on the fact that the difference between any two values of a Brownian motion process, at two different time instants, is always Normal-distributed random-variable with zero mean. For this Normal distributed difference, if 3-dB bandwidth is denoted by $\beta$ and the time between the instants is denoted by $\tau$, then the variance is $4\pi\beta\tau$ [26]; namely $\left[\phi(t) - \phi(t+\tau)\right] \sim \mathcal{N}(0, 4\pi\beta\tau)$, if $\phi(t)$ is Brownian motion with 3-dB bandwidth of $\beta$. The expression in (23) above is a general form for the quantity $\mathrm{E}\left[\left|J_k(b,\delta)\right|^2\right]$. Below, the two scenarios of independent oscillators or the common oscillator in the transmitter and the receiver paths are explored further.

We consider first the *independent oscillators case*. Then, the phase noise processes $\phi_t(.)$ and $\phi_r(.)$ are statistically independent of each other, and (23) can be rewritten as

$$
\begin{aligned}
P_k^{ind} &= \frac{1}{N^2} \sum_{n=0}^{N-1} \sum_{n'=0}^{N-1} e^{-4\left|n'-n\right|\pi\beta T_s + j2\pi k(n'-n)/N} \\
&= \frac{1}{N^2} \sum_{n=n-N}^{N-1} \left(N - \left|n\right|\right) e^{-4\left|n\right|\pi\beta T_s + j2\pi kn/N} \\
&= \frac{1}{N^2} \left[ -N + \sum_{n=0}^{N-1} 2\left(N-n\right) e^{-4\pi n\beta T_s} \cos\left(2\pi kn \,/\, N\right) \right].
\end{aligned}
\tag{24}
$$

The first form in (24) is based on the fact that the two phase-noise difference terms in (23), namely $\phi_t\left((n-b)T_s - \delta\right) - \phi_t\left((n'-b)T_s - \delta\right)$ and $\phi_r(n'T_s) - \phi_r(nT_s)$, are both Normal-distributed random-variables with zero-mean, and now mutually independent of each other. Therefore, the expectation in (23) yields just the combined variance of these two random variables. The second and final forms in (24) follow then from plain algebraic manipulations. The final form is deployed in (14). Q.E.D.

Next, we consider the *common oscillator scenario* between the transmitter and the receiver sides, and hence have $\phi_t(t) = \phi_r(t) = \phi(t)$. As a result, we can rewrite (23) as

$$
\begin{aligned}
P_k^{com} &= \frac{1}{N^2} \sum_{n=0}^{N-1} \sum_{n'=0}^{N-1} e^{-\frac{1}{2}\mathrm{E}\left[\left|\phi\left((n-b)T_s-\delta\right) - \phi\left((n'-b)T_s-\delta\right) + \phi(n'T_s) - \phi(nT_s)\right|^2\right] + j2\pi k(n'-n)/N} \\
&= \frac{1}{N^2}\left[ -N + \sum_{n=0}^{b} 2(N-n)e^{-4\pi n\beta T_s}\cos\left(2\pi kn\,/\,N\right) \right. \\
&\qquad \left. + \sum_{n=b+1}^{N-1} 2(N-n)e^{-4\pi b\beta T_s - 4\pi\beta\delta}\cos\left(2\pi kn\,/\,N\right) \right].
\end{aligned}
\tag{25}
$$

Examining the first line in (25) we can observe that the statistical properties of the random variable inside the expectation operator change as a function of the difference between the sum indices $n$ and $n'$. Therefore, in the analysis, the indices can be replaced by their separation when working towards the final expression. Then, to progress further requires somewhat burdensome but still straight-forward step-by-step manipulations where the well-known properties of the Wiener process are exploited. More specifically, one can separate the phase noise terms inside the expectation in a way that for every $n$, $n'$ pair for every value of $b$, the two phase-noise differences are grouped separately so that they are statistically independent, i.e., in a way that the two pairs of phase noise processes do not overlap in time. Then their combined variances can be calculated separately and finally summed together which results in the two sums written in the second and the final form of (25). Writing out these intermediate steps is skipped here, for compactness of presentation, and hence the second form of (25) represents the final result, deployed in (15). Q.E.D.

## BIOGRAPHIES

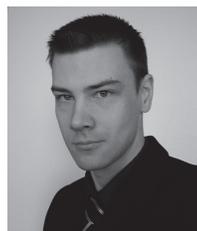

**Ville Syrjälä** (S'09, M'12) was born in Lapua, Finland, in 1982. He received the M.Sc. (Tech.) degree in 2007 and D.Sc. (Tech.) degree in 2012 in communications engineering (CS/EE) from Tampere University of Technology (TUT), Finland.

He was working as a research fellow with the Department of Electronics and Communications Engineering at TUT, Finland, until 2013. Currently, he is working as a postdoctoral fellow of Japan Society for the Promotion of Science (JSPS) at Kyoto University, Japan. His general research interests are in full-duplex radio technology, communications signal processing, transceiver impairments, signal processing algorithms for flexible radios, transceiver architectures, direct sampling radios, and multicarrier modulation techniques.

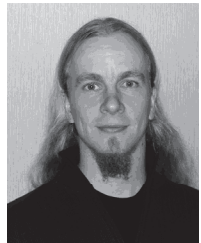

**Mikko Valkama** (S'00, M'02) was born in Pirkkala, Finland, on November 27, 1975. He received the M.Sc. and Ph.D. degrees (both with honours) in electrical engineering (EE) from Tampere University of Technology (TUT), Finland, in 2000 and 2001, respectively. In 2002 he received the Best Ph.D. Thesis award by the Finnish Academy of Science and Letters for his dissertation entitled "Advanced I/Q signal processing for wideband receivers: Models and algorithms".

In 2003, he was working as a visiting researcher with the Communications Systems and Signal Processing Institute at SDSU, San Diego, CA. Currently, he is a Full Professor and Department Vice Head at the Department of Electronics and Communications Engineering at TUT, Finland. He has been involved in organizing conferences, like the IEEE SPAWC'07 (Publications Chair) held in Helsinki, Finland. His general research interests include communications signal processing, estimation and detection techniques, signal processing algorithms for software defined flexible radios, full-duplex radio technology, cognitive radio, digital transmission techniques such as different variants of multicarrier modulation methods and OFDM, radio localization methods, and radio resource management for ad-hoc and mobile networks.

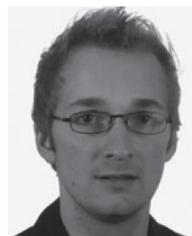

**Lauri Anttila** (S'06, M'11) received his Ph.D. (with honours) from Tampere University of Technology (TUT), Tampere, Finland, in 2011.

Currently, he is a Research Fellow at the Department of Electronics and Communications Engineering at TUT. His current research interests include statistical and adaptive signal processing for communications, digital front-end signal processing in flexible radio transceivers, and full-duplex radio systems.

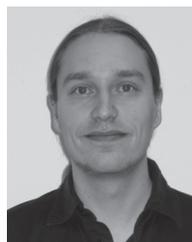

**Taneli Riihonen** (S'06) received the M.Sc. degree in communications engineering (with distinction) from Helsinki University of Technology, Helsinki, Finland in February 2006.

During the summer of 2005, he was an intern at Nokia Research Center, Helsinki, Finland. Since fall 2005, he has been a researcher at the Department of Signal Processing and Acoustics, Aalto University School of Electrical Engineering, Helsinki, Finland, where he is completing his D.Sc. (Tech.) degree in the near future. He has also been a student at the Graduate School in Electronics, Telecommunications and Automation (GETA) in 2006-2010. His research activity is focused on physical-layer OFDM(A), multiantenna and relaying techniques.

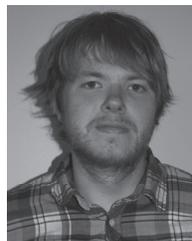

**Dani Korpi** was born in Ilmajoki, Finland, on November 16, 1989. He received the B.Sc. degree (with honors) in communications engineering from Tampere University of Technology (TUT), Tampere, Finland, in 2012, and is currently pursuing the M.Sc. degree in communications engineering at TUT.

In 2011, he was a Research Assistant with the Department of Signal Processing at TUT. Since 2012, he has been a Research Assistant with the Department of Electronics and Communications Engineering, TUT. His main research interest is the study and development of single-channel full-duplex radios, with a focus on analysing the RF impairments.